# Extremely high reflection of solar wind protons as neutral hydrogen atoms from regolith in space

**Martin Wieser[1*], Stas Barabash[1], Yoshifumi Futaana[1], Mats Holmström[1],**

**Anil Bhardwaj[2], R Sridharan[2], MB Dhanya[2],**

**Peter Wurz[3], Audrey Schaufelberger[3],**

**Kazushi Asamura[4]**

[1] Swedish Institute of Space Physics, Box 812, SE-98128 Kiruna, Sweden

* email: wieser@irf.se

[2] Space Physics Laboratory, Vikram Sarabhai Space Center, Trivandrum 695 022, India

[3] Physikalisches Institut, University of Bern, Sidlerstrasse 5, CH-3012 Bern, Switzerland

[4] Institute of Space and Astronautical Science, 3-1-1 Yoshinodai, Sagamihara, Japan

**Abstract**

We report on measurements of extremely high reflection rates of solar wind particles from regolith-covered lunar surfaces. Measurements by the Sub-keV Atom Reflecting Analyzer (SARA) instrument on the Indian Chandrayaan-1 spacecraft in orbit around the Moon show that up to 20% of the impinging solar wind protons are reflected from the lunar surface back to space as neutral hydrogen atoms. This finding, generally applicable to regolith covered

24     atmosphereless bodies, invalidates the widely-accepted assumption that regolith

25     almost completely absorbs the impinging solar wind.

26

27

28     **1. Introduction**

29     In the Solar System, space weathering results in most surfaces of atmosphereless bodies

30     being covered by regolith, a layer of loose, heterogeneous material of small grain size

31     (Clark et al., 2002). In the absence of an atmosphere, plasma interacts directly with the

32     surface, e.g. solar wind with the lunar regolith, and it has been tacitly assumed that the

33     plasma is almost completely absorbed (<1% reflected) in the surface material (e.g. Crider

34     et al. 2002, Schmitt et al., 2000, Feldman et al., 2000, Behrisch & Wittmaack, 1991).

35     Regolith reaches hydrogen saturation on a geologically very short timescale of at most

36     $10^4$ years (Johnson & Baragiola, 1991). Once saturated, for every impinging proton a

37     hydrogen atom is removed from the surface by sputtering, a scatter or desorption process.

38     Here we present measurements that invalidate the widely-accepted assumption that

39     regolith almost completely absorbs the impinging solar wind. A large fraction of up to

40     20% is reflected as energetic neutral hydrogen atoms back to space.

41

42     **2. Instrumentation**

43     The Sub-keV Atom Reflecting Analyzer (SARA) instrument (Barabash et al., 2009)

44     onboard the Indian Chandrayaan-1 spacecraft (Goswami & Annadurai, 2009) orbiting the

45     Moon in a 100-km polar orbit measures the neutral atom flux from the lunar surface and

46     simultaneously monitors the impinging flux of solar wind protons. The SARA instrument consists of two sensors: the Solar Wind Monitor (SWIM) (McCann et al., 2007) measures solar wind ions and the Chandrayaan-1 Energetic Neutrals Analyzer (CENA) (Kazama et al., 2007) measures energetic neutral atoms (10eV-3keV) arriving from the direction of the lunar surface. Both sensors provide angular and mass resolution: SWIM has a field-of-view of 7.5° x 180° divided into 16 angular pixels covering directions from nadir to zenith on the sun facing side of the spacecraft. The orbital motion of Chandrayaan-1 is used to scan the SWIM field-of -view across the full solar wind angular distribution. The center of the CENA field-of-view is nadir-pointing and in total 160° cross-track times 7° along-track in size, divided into 7 angular pixels. Mass resolution m/dm of both sensors is 2-3, depending on sensor, angular pixel and mass. The geometric factor for SWIM is 5 $\cdot 10^{-5}$ cm$^2$ sr eV/eV for 500 eV protons. The geometric factor of CENA is 5 $\cdot 10^{-8}$ cm$^2$ sr eV/eV for 500eV neutral hydrogen. Both values are for a single angular pixel at the center of the field-of-view.

## 3. Observations

During nominal solar wind conditions SARA observed that up to 16-20% of the proton flux impinging on the lunar surface is reflected back to space as energetic neutral hydrogen atoms. SARA persistently detects high intensity fluxes of energetic neutral hydrogen atoms propagating from the lunar surface while the latter is illuminated by the solar wind. The reflected neutral atom flux changes consistently with the change in solar wind incident angle (Figure 1), with the reported numbers reached at the lunar equator. No significant energetic neutral atom flux is observed on the night side. As the intensity

69    of the impinging solar wind proton flux changed orbit by orbit, the intensity of the

70    reflected neutral atom flux changed correspondingly and consistently (Figure 2). The

71    reflected neutral hydrogen that was observed has an energy spectrum with a distinct

72    upper energy of 300 eV, roughly half of the measured central energy (540 eV) of the

73    impinging protons. Upon reflection from the surface, solar wind particles experience an

74    average energy loss of more than 50%.

75

76    **4. Discussion and Conclusions**

77    Ions with solar wind energies interacting with surfaces are likely to be neutralized e.g. by

78    resonant or Auger neutralization, before being absorbed or reflected (e.g. Niehus et al.,

79    1993). We observe that up to 20% of solar wind protons are reflected from the lunar

80    surface back to space as energetic neutral hydrogen atoms with energies larger than 25

81    eV (Figure 2). This high reflection invalidates the previous assumption that regolith

82    almost completely absorbs the impinging solar wind. On a saturated surface the sum of

83    all loss processes equals the solar wind precipitation, the high reflection yield reduces

84    thus the amount of hydrogen available for release by other processes, such as sputtering

85    or desorption, both mechanisms predominantly producing neutrals with energies of only

86    a few eV (Betz & Wien, 1994). Since the energy of the reflected energetic neutral

87    hydrogen is much higher than the escape energy, escaping atoms propagate on ballistic

88    trajectories. They can be used for detailed remote imaging of the solar wind–surface

89    interaction (Bhardwaj et al., 2005, Futaana et al., 2006). The average energy loss of more

90    than 50% when interacting with the surface is likely due to several mechanisms. A simple

91    classical binary collision model (Niehus et al, 1993) gives an energy loss between 10%

92    and 20%. This indicates the presence of other loss mechanisms, such as multiple

93    scattering processes or interaction with the crystal lattice. Another possible loss process is

94    retarding of impinging solar wind ions by a positive surface potential on the dayside

95    (Vondrak, 1992), before the ions interact with the regolith surface. The observed lunar

96    energetic neutral hydrogen component with an energy of up to 300eV should also be

97    visible in measurements of Doppler-shifted Lyman alpha emissions (Gott & Potter,

98    1970). The rate of implantation of solar wind $^4$He and $^3$He in the lunar regolith has to be

99    reconsidered as well, because it can be expected that helium will have a similar reflection

100   rate to hydrogen. Helium ions are efficient in creating hydrogen recoils when impinging

101   on the surface (Niehus et al., 1993), a process that will additionally remove hydrogen

102   from the lunar surface. It is expected that high solar wind reflection occurs on similar

103   bodies in the Solar System as well, e.g. on Mercury, planetary satellites such as Phobos or

104   Deimos, or on asteroids. Finally, the high neutral hydrogen reflection implies a lower

105   hydrogen implantation rate in the regolith. Earlier work on the hydrogen influx on the

106   lunar surface assumed almost complete absorption of solar wind ions (Crider et al. 2002,

107   Schmitt et al., 2000, Feldman et al., 2000, Behrisch & Wittmaack, 1991), a finding which

108   has to be revisited in the light of the present results: in equilibrium conditions, escaping

109   and trapped hydrogen must equal the impinging hydrogen from solar wind. The efficient

110   reflection process reduces the possible trapping rate e.g. at the lunar poles. This makes it

111   less likely that hydrogen at the poles is entirely of solar wind origin.

112

113

114

**Figure 1**: Neutral hydrogen measurements recorded during three consecutive orbits on 6 February 2009. The flux from the surface seen with the nadir-looking sensor pixel is shown by thin colored zenith-pointing lines. Data from all three orbits covering an area around 8° W lunar longitude on the lunar dayside are superimposed. Color and the line length depict the flux intensity. Solar wind ions impinge onto the lunar surface approximately from the Sun direction indicated by the green axis. A clear dependence of the flux on solar zenith angle is seen on the dayside. Small fluxes on the night side correspond to the instrument background. The lunar surface map is taken from Clementine image data.

**Figure 2**: Energy spectra of the solar wind (right side, open squares) and of the corresponding reflected energetic hydrogen (left side, open circles). Data from the same three orbits on 6 February 2009 as used in Figure 1 are shown, with dayside equator crossings at 05:22 UTC (solid lines), 07.20 UTC (dashed lines) and 09:18 UTC (dotted lines). Measurements were taken shortly after the Moon crossed the Earth's bow shock to the downstream direction, resulting in a broadening of the energy spectrum of the solar wind protons. Note the good correlation between the reflected energetic neutral flux and the solar wind flux variations. The reflection yield when crossing the equator is 0.16 to 0.20, depending on the orbit. Taking uncertainties arising from detector efficiency into account and assuming an isotropic angular scatter distribution of the energetic hydrogen atoms, the average values over several orbits are, with 95% confidence, between 0.03 and 0.35. The isotropic angular scattering distribution on the surface assumed for calculating the total reflected flux is based on data obtained from pixels looking in non-nadir

188    directions. Error bars shown represent knowledge of the instrument geometric factor (1σ).

189

190

191

192

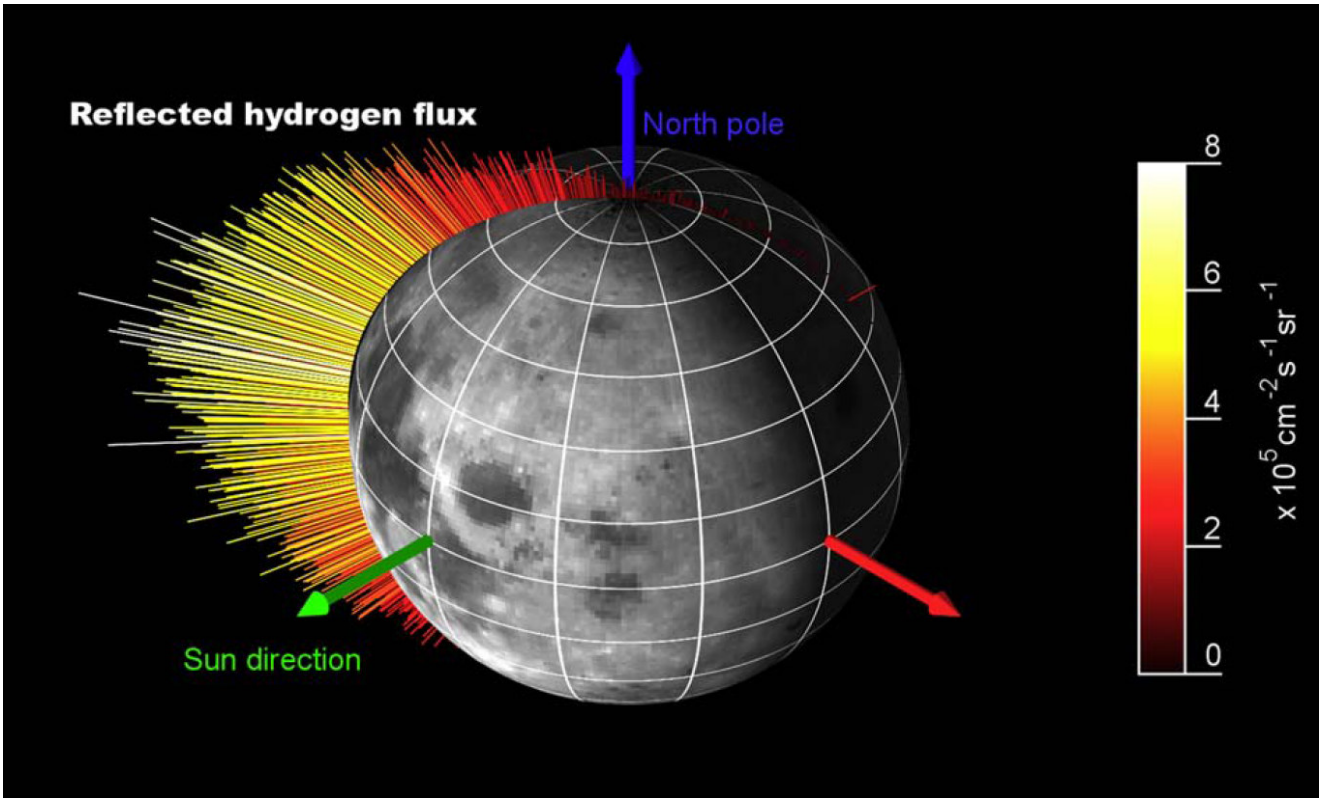

Figure 1

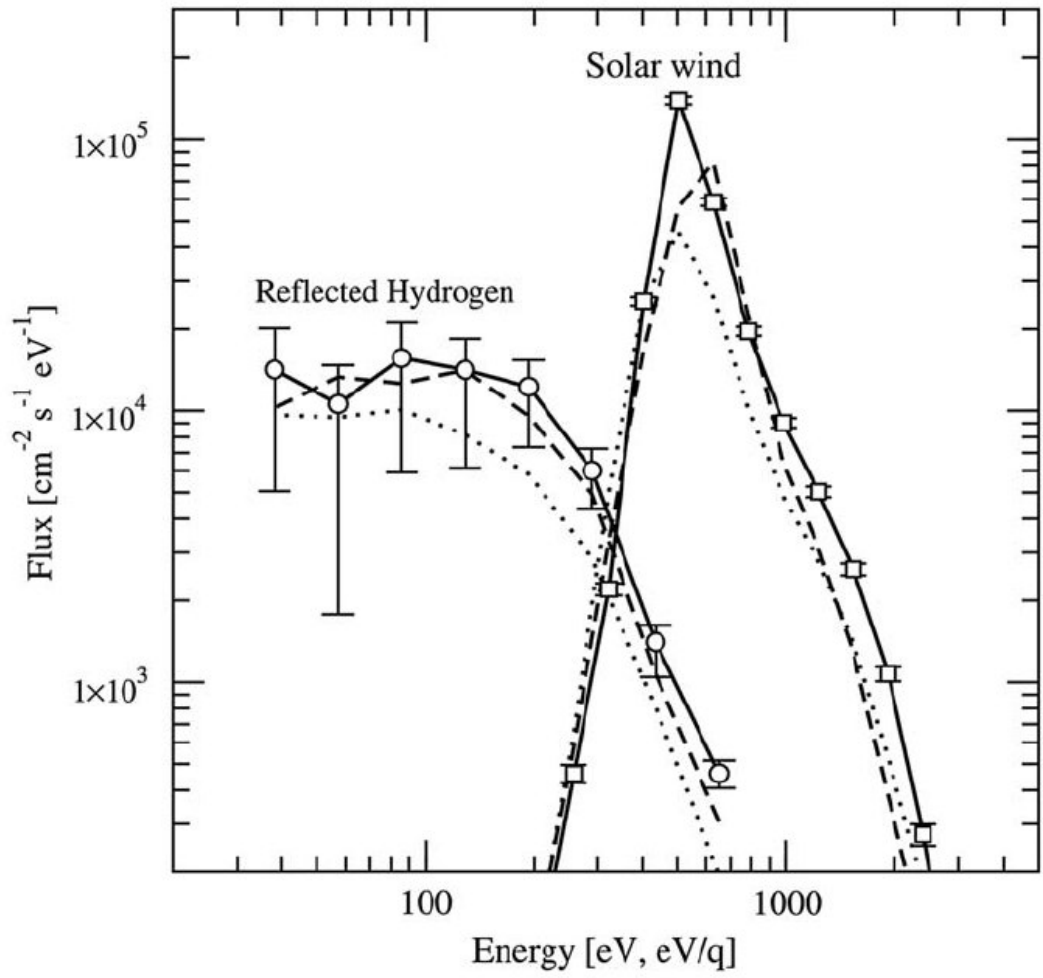

Figure 2